\documentclass{article}

\usepackage{amsmath,amsfonts,latexsym,amssymb}
\usepackage{verbatim,amsthm,graphicx}
\usepackage{wrapft}

\def\FigDir{.}
\def\abstract#1{\vskip 7mm 
        \begin{center}{\large Abstract}\par \smallskip
                \begin{minipage}[c]{12cm}
                        \small #1
                \end{minipage}
        \end{center}
}
\def\title#1{\begin{center}{\Large\bf #1}\end{center}}
\def\author#1{\vskip 5mm \begin{center}{#1}\end{center}}
\def\address#1{\begin{center}{\it #1}\end{center}}

\usepackage{amsmath}
\usepackage{amssymb}
\usepackage{mathrsfs}

\def\ifempty#1{\def\tmpdata{#1}\ifx\tmpdata\empty }

\def\linebreak{\hfill\break}

\def\bra<#1|{\langle #1\rvert}
\def\ket|#1>{\lvert#1 \rangle}
\def\braket<#1|#2>{\langle #1|#2 \rangle}

\def\pfrac#1#2{\left(\frac{#1}{#2}\right)}
\def\const{\text{const}}
\def\otop#1{\hbox{$#1\kern-0.1em$\llap{\hbox{\raise1.7ex\hbox{$\scriptstyle\circ$}}}} }

\def\inpare#1{\left(#1\right)}
\def\bigpare(#1){\left(#1\right)}

\def\insbra#1{\left[ #1 \right]}

\def\bigbra[#1]{\left[ #1 \right]}

\def\t{\tilde }

\def\therefore{\mbox{\setbox0=\hbox{X}\hbox{$\ldotp$}\raise0.7\ht0\hbox{$\ldotp$}\hbox{$\ldotp$}} \quad }
\def\because{\mbox{\setbox0=\hbox{X}\raise0.7\ht0\hbox{$\ldotp$}\hbox{$\ldotp$}\raise0.7\ht0\hbox{$\ldotp$}}\kern0pt }

\def\ZR{{{\mathbb Z}}}

\def\UG{{\rm U}}

\def\SU{{\rm SU}}

\def\upin{\hbox{\setbox0=\hbox{$\cup$} \vrule width 0.05 \wd0 height \ht0 depth 0pt \kern - 0.5\wd0 \box0 }}
\def\Frac(#1/#2){\left(\frac{#1}{#2}\right)}

\def\sdprod{\mathrel{{\setbox0=\hbox{$\displaystyle\times$}\lower0.3\wd0\hbox{$\stackrel{\box0}{\scriptstyle\sim}$}}}}

\def\tosigma#1,{%
    \ifx\tmpindex\relax \def\tmpindex{#1} \let\next=\tosigma
    \else \ifnum\tmpindex=0 1 \else \sigma_\tmpindex \fi
          \ifx#1\relax  \let\next=\relax
          \else \otimes \let\next=\tosigma \def\tmpindex{#1} \fi
    \fi \next}
\def\tspb(#1){\let\tmpindex=\relax\tosigma#1,\relax,}
\def\pd{\partial}

\def\Eq#1{\begin{equation} #1 \end{equation}}

\def\Eqr#1{\begin{eqnarray} #1 \end{eqnarray}}

\def\Eqrsubl#1#2{\begin{subequations}
  \expandafter\ifx\csname Rlabel\endcsname \relax \label{#1}
  \else \Rlabel{#1} \fi \Eqr{#2}\end{subequations}}
\def\Bitm{\begin{itemize}}
\def\Eitm{\end{itemize}}
\def\Blist#1#2{\begin{list}{#1}{\parsep=0pt \itemsep=0pt%
  \listparindent=0pt #2}}
\def\Elist{\end{list}}
\long\def\ignore#1#2{\def\ignoreflag{#1}\long\def\tmptext{#2}
  \ifnum\ignoreflag>1 #2 \fi}

\begin{document}
\title{\large Repulsons in the 5D Myers-Perry Family}

\author{%
Hideo Kodama%
\footnote{E-mail: Hideo.Kodama@kek.jp}
}
\address{%
IPNS, KEK and the Graduate University of Advanced Studies,\\
1-1 Oho, Tsukuba 305-0801, Japan
}

\abstract{
In this talk, we point out that curvature-regular asymptotically flat solitons with negative mass are contained in the Myers-Perry family in five dimensions. These solitons do not have horizon, but instead a conical NUT singularity of quasi-regular nature surrounded by naked CTCs. We show that this quasi-regular singularity can be made regular for a set of discrete values of angular momentum by introducing some periodic identifications, at least in the case in which two angular momentum parameters are equal. Although the spatial infinity of the solitons is diffeomorphic to $S^1\times S^3/\ZR_n$ ($n\ge3$), the corresponding spacetime is simply connected and asymptotically flat.
}


\section{Introduction}

\begin{figure}[b]
\begin{minipage}{11cm}
\centerline{
\includegraphics*[height=5cm]{\FigDir/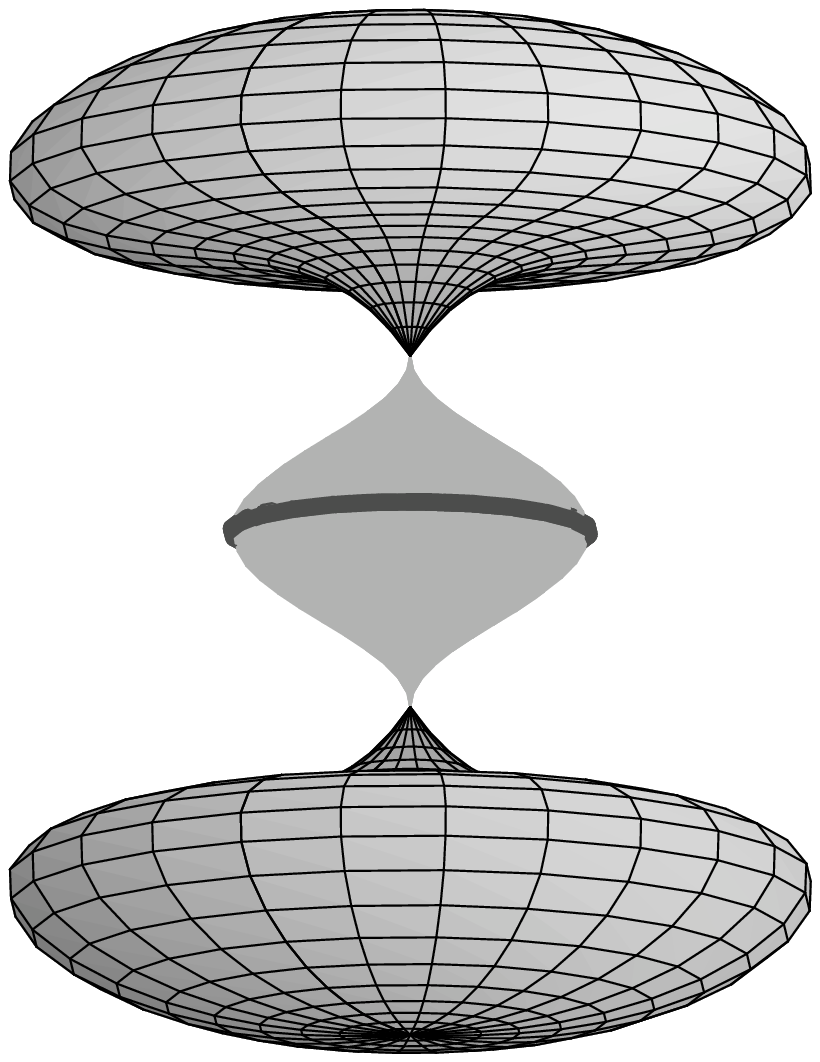}\hspace{2cm}
\includegraphics*[height=4cm]{\FigDir/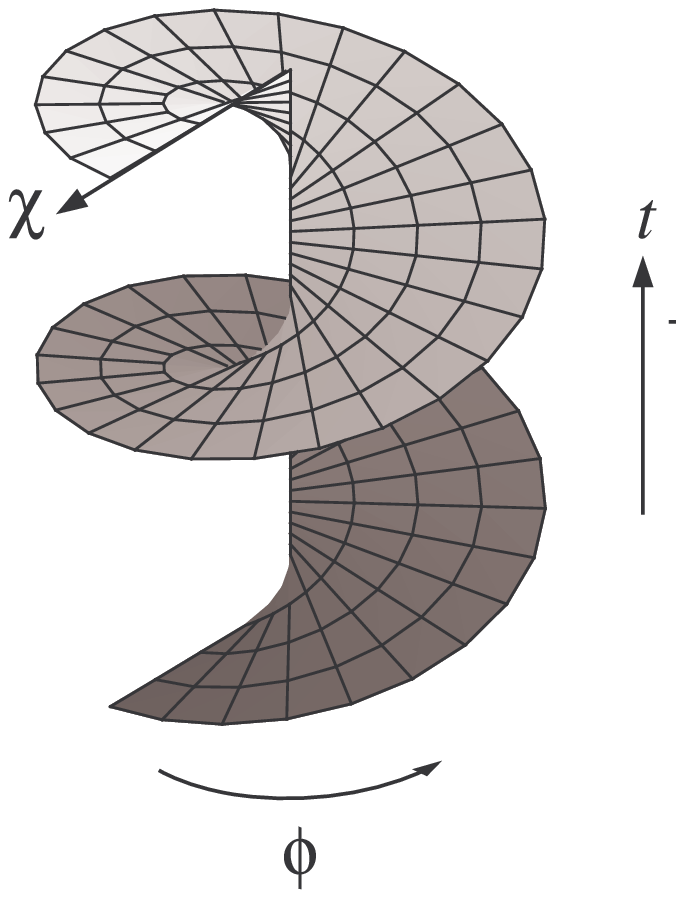}
}
\end{minipage}
\begin{minipage}{4cm}
\noindent 
\caption{\label{fig:TS2}TS2 solution.}

\small The left panel show the two horizons connected by a region where the angular Killing vector becomes time-like and the ring singularity on its boundary. The right panel show the helical structure of the time coordinate on the $z$-axis connecting the two horizons\cite{Kodama.H&Hikida2003}.
\end{minipage}
\end{figure}

In four dimensions, there are some asymptotically flat regular black hole solutions with CTCs inside horizon. However, no asymptotically flat regular simply-connected vacuum solution with naked CTCs has been found so far. For example, the Kerr and the Kerr-Newman solutions have  CTCs, but they become naked only in the superextreme case with naked ring singularity. Similarly, the Tomimatsu-Sato solution with $\delta=2$\cite{Tomimatsu.A&Sato1972} has double horizons and circles generated by the rotational Killing vector become time-like near the $z$-axis segment connecting two horizons, but the spacetime has the well-know naked ring singularity and a conical singularity along the segment\cite{Kodama.H&Hikida2003}(see Fig. \ref{fig:TS2}). Here, the asymptotically flatness is a crucial condition, because we have the Taub-NUT solution and the G\"odel solution as famous examples of curvature-regular vacuum solutions that have CTCs but are not asymptotically flat.

In higher dimensions, the situation changes. For example, the BMPV solution\cite{Breckenridge.J&&1997,Kallosh.R&Rajaraman&Wong1997} in a five-dimensional gauged supergravity theory represents a rotating charged BPS black hole for slow rotations, while when its angular velocity exceeds some critical value, the horizon turns to a surface called a repulson which is surrounded by CTCs and cannot be penetrated\cite{Gibbons.G&Herdeiro1999}. Similar BPS solutions were found in other supergravity theories in five dimensions\cite{Cvetic.M&&2005A,Gauntlett.J&&2003}.

These BPS solutions show that the existence of CTCs and  regularity and asymptotically flatness can coexist in higher dimensions. The main purpose of this talk is to point out that another set of examples of such spacetimes are contained in the five dimensional Myers-Perry solution\cite{Myers.R&Perry1986}. They are asymptotically flat regular spacetimes with negative mass and naked CTCs.

\section{Internal Structure of the 5D Myers-Perry Solution}

\subsection{Metric and regularity}

In terms of the Boyer-Lindquist coordinates $(t,{\phi_1},{\phi_2},r,\theta)$, the 5D Myers-Perry solution can be written as\cite{Myers.R&Perry1986}
\Eqr{
ds^2 &=& \frac{r^2\rho^2}{\Delta} dr^2 + \rho^2d\theta^2
  +(r^2+a^2)\sin^2\theta d{\phi_1}^2+ (r^2+b^2)\cos^2\theta d{\phi_2}^2
   \notag\\
   && -dt^2 + \frac{\mu }{\rho^2}\insbra{dt+a\sin^2\theta d{\phi_1} + b\cos^2\theta d{\phi_2}}^2,
\label{metric:5DMP:BLcoord}
}
where $\Delta :=(r^2+a^2)(r^2+b^2)- \mu  r^2$.

Because the determinant of the metric is given by $-g = r^2\rho^4 \sin^2\theta\cos^2\theta$, apart from the angular coordinate singularities at $\theta=0,\pi/2$, 
the metric \eqref{metric:5DMP:BLcoord} is apparently singular at $r=0$ and at $\Delta=0$ in addition to  $\rho=0$, which is curvature singularity from the expression for the Kretchman invariant,
\Eq{
R_{\mu\nu\lambda\sigma}R^{\mu\nu\lambda\sigma}
=24 \mu ^2 \frac{(3r^2-a^2\cos^2\theta-b^2\sin^2\theta)(r^2-3a^2\cos^2\theta-3b^2\sin^2\theta)}{\rho^{12}}.
}

Among these apparent singularities, we can confirm that the points with $r=0$ are not real singularity by expressing the metric in terms of $x=r^2$ as\cite{Myers.R&Perry1986}
\Eqr{
ds^2 &=& \frac{\rho^2}{4\Delta} dx^2 + \rho^2d\theta^2
    +(x+a^2)\sin^2\theta d{\phi_1}^2+ (x+b^2)\cos^2\theta d{\phi_2}^2
   \notag\\
   && -dt^2 + \frac{\mu }{\rho^2}\insbra{dt+a\sin^2\theta d{\phi_1} + b\cos^2\theta d{\phi_2}}^2.
}
Now, $\Delta$ and $\rho^2$ become polynomials in $x$ as
\Eq{
\Delta=(x+a^2)(x+b^2)-\mu x,\quad
\rho^2=x+a^2\cos^2\theta+b^2\sin^2\theta,
}
and $-g$ is expressed as $-g=\rho^4 \sin^2\theta \cos^2\theta/4$. 

\subsection{Structure of Killing orbits}


On a Killing horizon, the induced metric on the orbit space of the Killing vectors $\pd_t, \pd_{\phi_1}$ and $\pd_{\phi_2}$ becomes degenerate. From $\left|g_{a b}\right|_{a,b=t,{\phi_2},{\phi_1}}=-\Delta \sin^2\theta\cos^2\theta$, this condition can be written 
$\Delta=0$. When $ab\neq0$ and $\mu>0$, this equation for $x$ has a root in the region $\rho^2>0$ iff $\mu >(|a|+|b|)^2$. In this case, there actually exist two positive roots, which are both regular horizons. 

In contrast, when  $\mu <0$, we have only negative roots, but one of them, $x=x_h$, is larger than $-b^2$ and in the region with $\rho^2>0$. This apparently indicates that even in the negative mass case, the spacetime has a regular horizon and a regular domain of outer communication. However, this is not the case. In fact, the spacetime has pathological features around $x=x_h$, when $x_h\neq0$. For example, the Killing vector $k$ whose norm vanishes at $x=x_h$ becomes spacelike outside the horizon. Further, the norm of the Killing vector $\pd_t$ is negative definite for $x\ge x_h$.

\subsection{CTCs}

Let $\Phi$ be the symmetric matrix of degree 2 defined by $\Phi:= \inpare{g_{ij}}_{i,j={\phi_1}.{\phi_2}}$. Then, we can show that $\det \Phi= D_1\sin^2\theta\cos^2\theta/\rho^2$ is non-negative in $x>0$ for $\mu >0$, and therefore that there occurs no causality violation at or outside the outer horizon for $\mu >(|a|+|b|)^2$.

In contrast, for $\mu <0$, causality violation occurs in the region  $x\ge x_h$. In fact, $D_1$ can be written as
\Eq{
D_1=\Delta(x+a^2\sin^2\theta+b^2\cos^2\theta+\mu ) + \mu ^2 x.
}
From this, around $x=x_h$ at which $\Delta=0$, the sign of $D_1$ is determined by the sign of $x_h$. Hence, when $x_h<0$, 2-tori generated by $\pd_{\phi_1}$ and $\pd_{\phi_2}$ become timelike surfaces around $x=x_h$, which always contain CTCs. As we show in the next section, this feature together with the fact that $g_{tt}$ is netative and $\Phi$ has a finite non-degenerate limit at $x=x_h$ implies that $x=x_h$ is not a horizon but rather something like a repulson, which was first found by Gibbons and Herdeiro for supersymmetric rotating black holes in five dimensions\cite{Gibbons.G&Herdeiro1999,Cvetic.M&&2005A}.

\section{Repulson}

\subsection{$\UG(2)$ MP solution}

In the case $a=b\neq0$, the metric in the Boyer-Lindquist coordinates reads
\Eq{
ds^2 = \frac{y dy^2}{4(y^2-\mu y+a^2\mu )}+yds^2(S^3)
 -dt^2+ \frac{\mu }{y}\inpare{dt+a\chi^3}^2,
}
where $y=\rho^2=x+a^2$, and $\chi^3$ is the 1-form on $S^3$ invariant under the action of $\SU(2)$
\Eq{
\chi^3=\sin^2\theta d{\phi_1} + \cos^2\theta d{\phi_2}.
}

\subsection{Negative mass soliton}

Now, we show that the surface corresponding to the root $x=x_h (>-a^2)$ of $\Delta=0$ for $\mu<0$, which is intrinsically $S^2$, is not a regular submanifold of the spacetime, but rather a quasi-regular singularity. For that purpose, let us introduce the coordinate 
\Eq{
\xi= \inpare{\frac{a^2-y_h}{2a^2-y_h}(y-y_h)}^{1/2},
}
where $y_h=x_h+a^2>0$. Then, in terms of the 1-forms
\Eq{
\t \tau= -\frac{a}{y_h} dt + \chi_3,\quad
\t \chi =\chi_3 + \frac{y_h}{2a^2-y_h} \t \tau,
}
the metric can be written near the locus $y=y_h$ as
\Eq{
ds^2 \simeq -\frac{y_h^2}{(a^2-y_h)}\t \tau^2
    + d\xi^2  + c^2\xi^2 \t\chi_3^2 + y_h  ds^2(S^2),
\label{metric:aroundrepulson}
}
where
\Eq{
c= \frac{2a}{(a^2-y_h)^{1/2}} \inpare{1-\frac{y_h}{2a^2}}.
}

The curvature tensor of this metric is bounded everywhere. However, it has a kind of conical singularity at $\xi=0$ because $c>2$ and $t$ is not a periodic coordinate. This conical singularity cannot be removed even by a periodic identification of $t$ in general. For some discrete values of $y_h$, however, the spacetime can be made regular by such an identification. To see this, we utilise the local charts, $(z_+,w_+,t_+)$ for $\theta\neq\pi/2$ and $(z_-,w_-,t_-)$ for $\theta\neq0$ defined by
\Eq{
z_\pm:= (\tan\theta)^{\pm1}e^{\pm i \phi},\quad
w_\pm:= \xi e^{ic(\psi\pm\phi)/2},\quad
t_\pm:= t-\frac{y_h}{4a}(\psi\pm\phi).
}
It is easy to see that in each of these coordinate system, the metric \eqref{metric:aroundrepulson} has regular expressions.

In the region $0<\theta<\pi/2$ covered by both charts, the two set of coordinates are related by
\Eq{
z_-=\frac{1}{z_+},\quad
w_-= w_+ \pfrac{z_+}{|z_+|}^{-c},\quad
t_-= t_+ + i\frac{y_h}{a} \ln \frac{z_+}{|z_+|}.
}
From the relation for $w_\pm$, we find that $c$ must be an integer $n>2$ for the coordinate transformation to be well-defined. Further, because the $\ln$-term in the relation for $t_\pm$ produces ambiguity in the values of $t_\pm$ written as an integer multiple of $2\pi y_h/a$, these two charts are consistent only when we identify $t_\pm$ periodically with a period $2\pi y_h/(ma)$ where $m$ is some positive integer.

For $\xi\neq0$, by this identification, each $\xi=\const$ surface becomes a $S^1$ bundle over $S^3/\ZR_n$. However, since we can easily show that the regular manifold constructed above from the region around $\xi=0$ is simply connected for $m=1$, the whole spacetime is also simply connected for $m=1$. In that case, CTCs along $S^1$ produced by the above identification at $\xi\neq0$ cannot be removed by taking a covering space.

\section{Summary and Discussion}

In the present paper, we have shown that the five-dimensional Myers-Perry solution with negative mass describes an asymptotically flat rotating spacetime with naked CTCs, no horizon and no curvature singularity. We further pointed out that we can construct a regular repulson-type asymptotically flat soliton from them for a discrete set of values of the angular momentum for a fixed mass parameter $\mu$. 

  Although the structure of the original metric when the repulson appears is similar to that of BPS solutions with repulsons, there exist several big differences between our family and the BPS families. In particular, our solutions are non-BPS solution to the purely vacuum Einstein equations, and the repulson is a regular submanifold of the whole spacetime. This should be contrasted to the BPS case in which the repulson is at spatially infinite distance. Further, the soliton spacetime obtained after regularisation is not asymptotically flat in the standard sense because it has some time periodicity at spatial infinity and its spatial infinity is $S^{3}/\ZR_n$ with $n\ge3$, although it is topologically simply connected. Finally, the repulson appears only for negative mass. This is rather striking because a regular soliton spacetime with negative mass can exist in higher dimensions. 

  Finally, we note that our analysis can be extended to the $\UG(N)$ symmetry Myers-Perry solutions in $2N+1$ dimensions\cite{Gibbons.G&Kodama2009A}. It can be also extended to the rotating black-hole type solutions with cosmological constant\cite{Gibbons.G&&2005} and to the black holes with generic angular momentum parameters. These extensions are under investigation. 

\section*{Acknowledgement}

This talk is based on the work in collaboration with G.W. Gibbons. The author would like to thank Hideki Maeda, Akihiro Ishibashi, Shinya Tomizawa and Andres Anabalon for discussions and useful comments. This work was supported in part by Grants-in-Aids for Scientific Research from JSPS (No. 18540265) and for the Japan-U.K. Research Cooperative Program.


\end{document}